 \documentclass[final,5p,times,twocolumn]{elsarticle}

\usepackage{rotating}
\usepackage{amsmath}

\usepackage{amssymb}

\journal{Nuclear Instruments and Methods in Physics Research A }

\begin{document}

\begin{frontmatter}



\title{Image Reconstruction with a LaBr$_3$-based Rotational Modulator}


\author{B. Budden}
\author{G.L. Case}
\author{M.L. Cherry}

\address{Department of Physics \& Astronomy, Louisiana State University, Baton Rouge, LA, USA 70803}

\begin{abstract}
A rotational modulator (RM) gamma-ray imager is capable of obtaining significantly better angular resolution than the fundamental geometric resolution defined by the ratio of detector diameter to mask-detector separation. An RM imager consisting of a single grid of absorbing slats rotating ahead of an array of a small number of position-insensitive detectors has the advantage of fewer detector elements (i.e., detector plane pixels) than required by a coded aperture imaging system with comparable angular resolution. The RM therefore offers the possibility of a major reduction in instrument complexity, cost, and power. A novel image reconstruction technique makes it possible to deconvolve the raw images, remove sidelobes, reduce the effects of noise, and provide resolving power a factor of 6 -- 8 times better than the geometric resolution. A 19-channel prototype RM developed in our laboratory at Louisiana State University features 13.8$^{\circ}$ full-angle field of view, 1.9$^{\circ}$ geometric angular resolution, and the capability of resolving sources to within 35$'$ separation. We describe the technique, demonstrate the measured performance of the prototype instrument, and describe the prospects for applying the technique to either a high-sensitivity standoff gamma-ray imaging detector or a satellite- or balloon-borne gamma-ray astronomy telescope.
	
\end{abstract}


\end{frontmatter}


\section{Introduction}

Imaging hard x-ray and gamma-ray photons is not possible using traditional focusing techniques. Instead, an indirect technique must be used, such as spatial/temporal modulation or Compton scattering. A single-grid Rotational Modulator (RM) is one such instrument, which combines the advantages of two more-commonly used imagers: the spatially-modulating coded aperture and the temporally-modulating double-grid Rotating Modulation Collimator (RMC).

\begin{figure}
	\begin{center}
	\begin{tabular}{c c c}
		\includegraphics[height=1.15in]{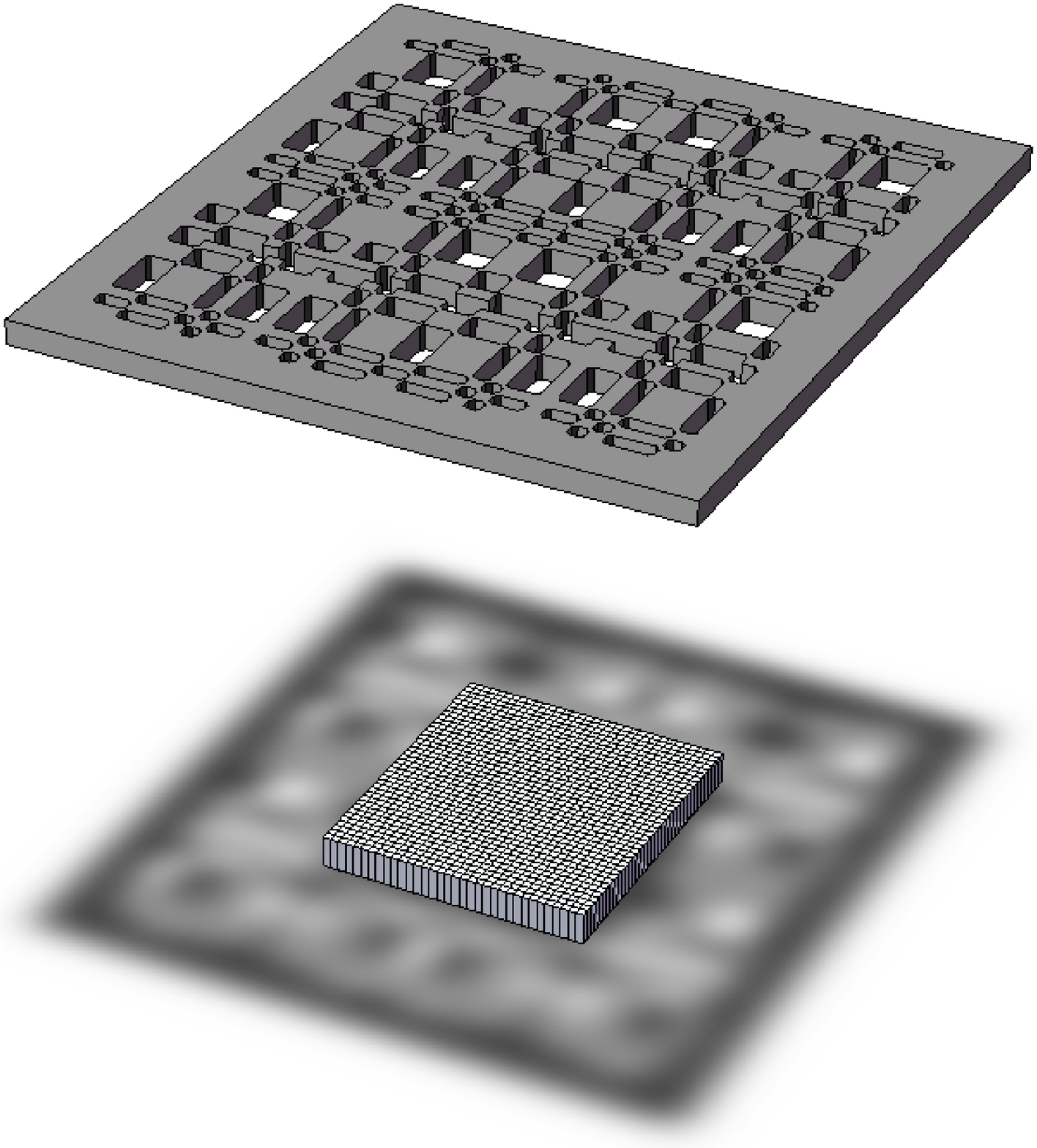} & \includegraphics[height=1.15in]{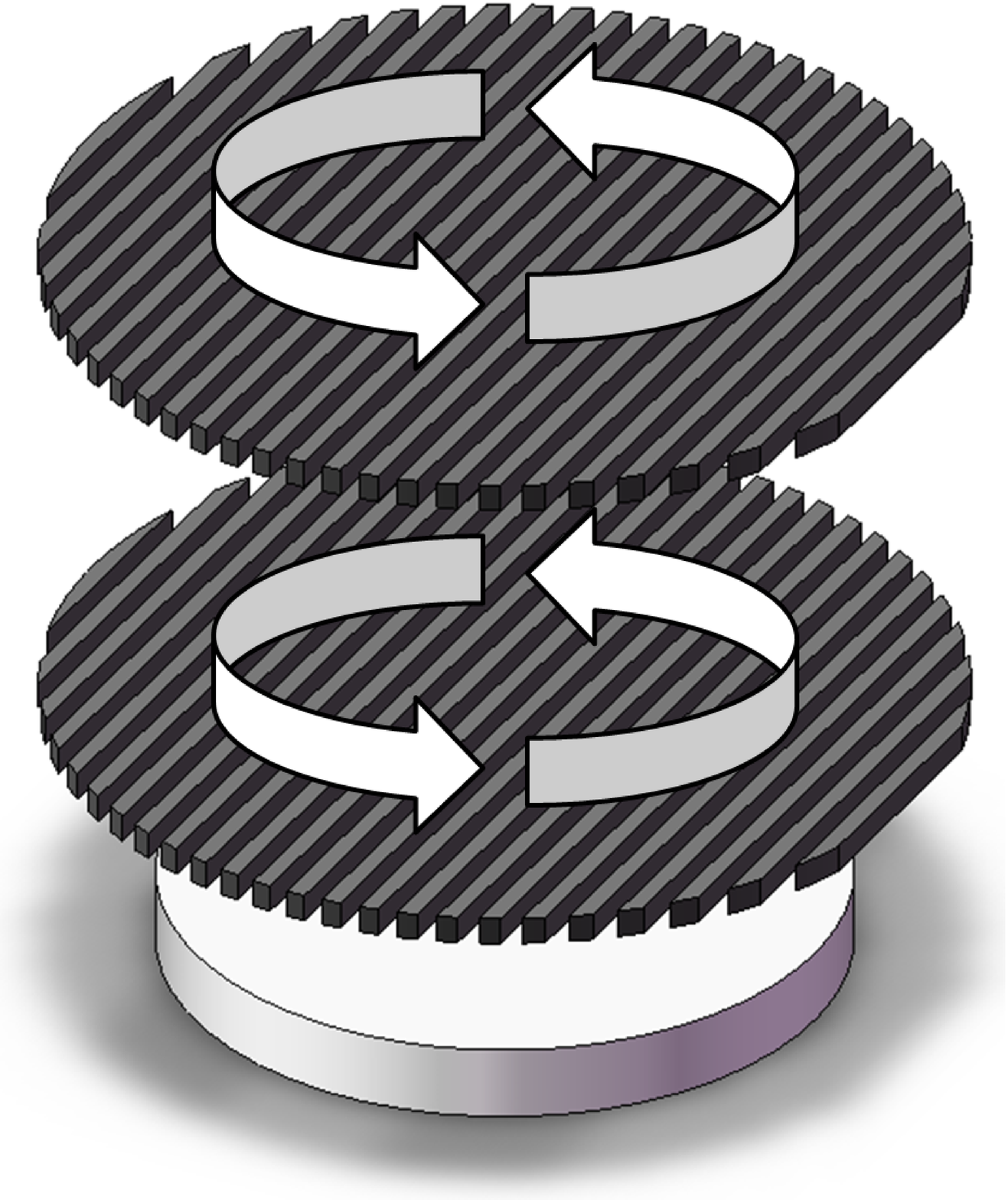} & \includegraphics[height=1.15in]{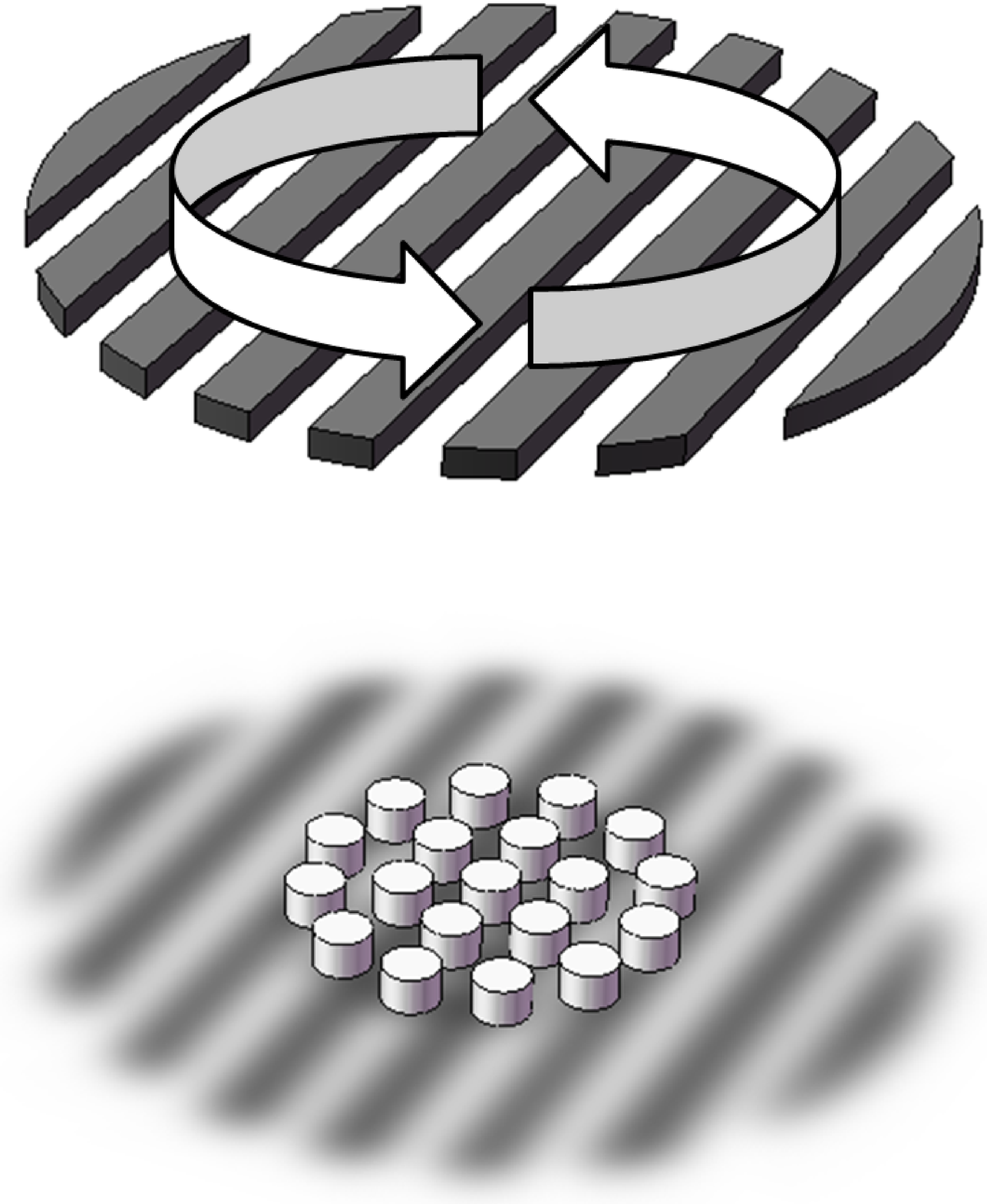} \\
		Coded Aperture & RMC & RM
	\end{tabular}
	\end{center}
	\caption[example]
   { \label{fig:multiplexers} 
   Basic structure of three multiplexing imagers. All instruments consist of a detection plane situated below a suspended mask which spatially or temporally modulates incident photons. The RM shares its rotating grid in common with the RMC, and its single-grid and multiple detector elements with the coded aperture.}
\end{figure}

A coded aperture (Fig. \ref{fig:multiplexers}, left) \cite{Caroli1987, Dicke1968, FenimoreCannon1978} consists of a mask of opaque and transparent pixels (typically equal numbers of each, resulting in a 50\% ``throughput'') arranged in a specific pattern. The source distribution in the object scene casts a shadow of the mask onto a position-sensitive detection plane. Angular resolution of the instrument is given by the ratio of mask pixel size to mask-detection plane separation, and can be made very good in practice. However, the size of the detection plane elements must be small enough to allow for adequate measurement of the shadow (less than or equal to half the mask pixel size in order to satisfy the Nyquist criterion). For a high-sensitivity, large-area instrument, this requirement results in a high number of readout channels, increasing cost and complexity. 

A Rotational Modulation Collimator (Fig. \ref{fig:multiplexers}, middle) \cite{Mertz1967, Schnopper1968} is one of a class of temporal modulators. It consists of two concentric offset grids of slats above a single position-insensitive detector. As the grids rotate in tandem, photons from off-axis sources will be modulated before detection. The detector measures a time history of counts, unique to the source distribution in the object scene. The angular resolution of the RMC, like the coded aperture, is defined by the width of the slat spacing divided by the mask-detector separation. Since the slat spacing can easily be made small, the RMC is capable of excellent angular resolution, despite having only a single readout channel. The secondary grid, however, reduces the sensitivity of the instrument (25\% throughput) and increases the overall weight.

A Rotational Modulator (Fig. \ref{fig:multiplexers}, right) \cite{Durouchoux1983, Dadurkevicius1985} is another temporal modulation imager, but consists of only a single grid of opaque slats suspended above a small array of detectors. For a large-area gamma-ray detector, where the grid must be thick and correspondingly massive, the use of one rather than two grids can potentially be a significant advantage. The slat widths, slat spacing, and detector diameters are equal. The grid is rotated, periodically blocking photons incident on the instrument. This time-dependent shadow is recorded by the detectors as a history of counts during an entire exposure and then folded modulo the grid rotational period to produce a single count profile for each detector. With temporal modulation and modest detection plane spatial resolution, the RM shares the advantages of both the coded aperture and the RMC. It maintains high sensitivity with a 50\% throughput, low weight, and has a relatively simple readout system. Again, the intrinsic geometric angular resolution is defined by the ratio of slat spacing to mask-detector separation. Since this spacing must be equal to the detector diameters, this geometric resolution is effectively limited. Thus, for the advantages of the RM to fully outweigh those of other multiplexing instruments, the reconstruction technique must be capable of resolving images beyond the geometric resolution, i.e. it must achieve ``super-resolution.''

\begin{figure}
	\begin{center}
	\includegraphics[width=3.0in]{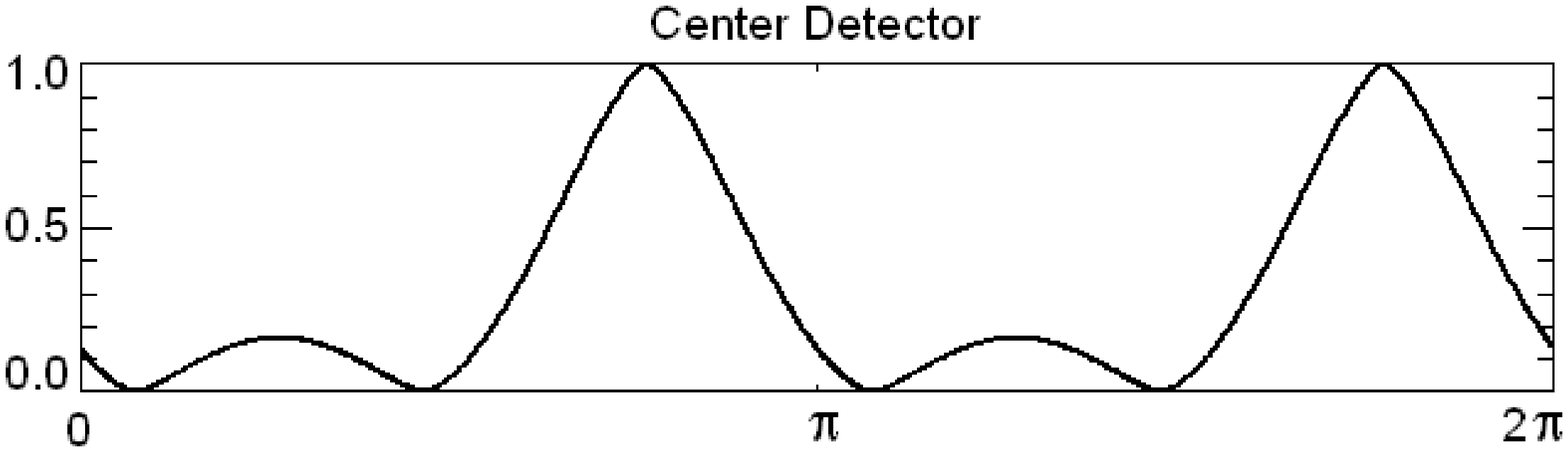} \\
	\includegraphics[width=3.0in]{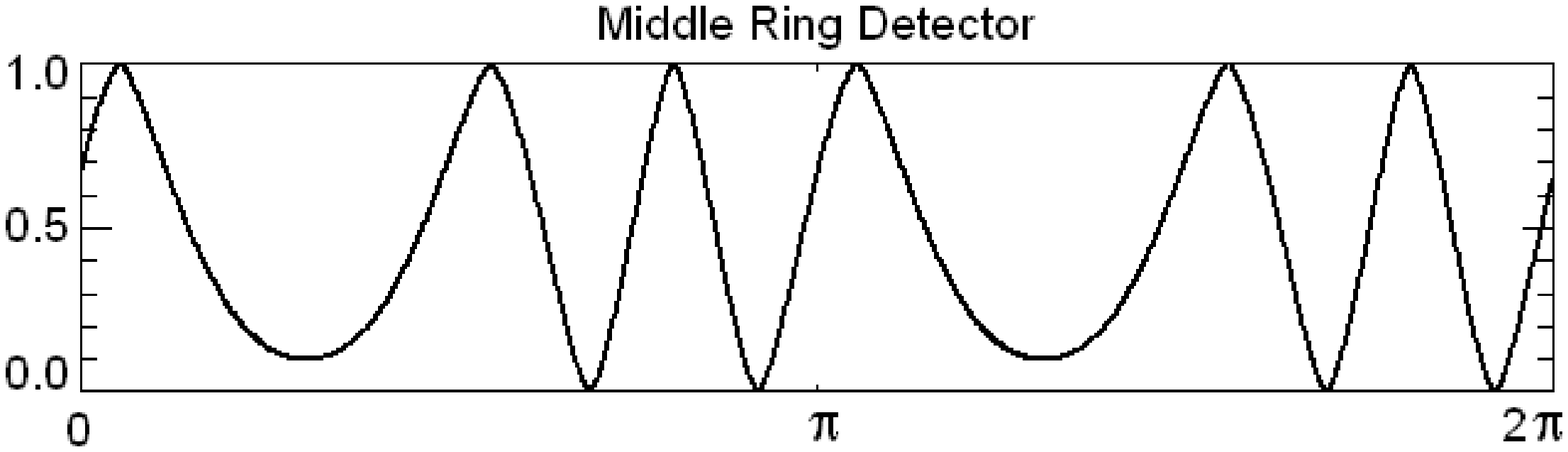} \\
	\includegraphics[width=3.0in]{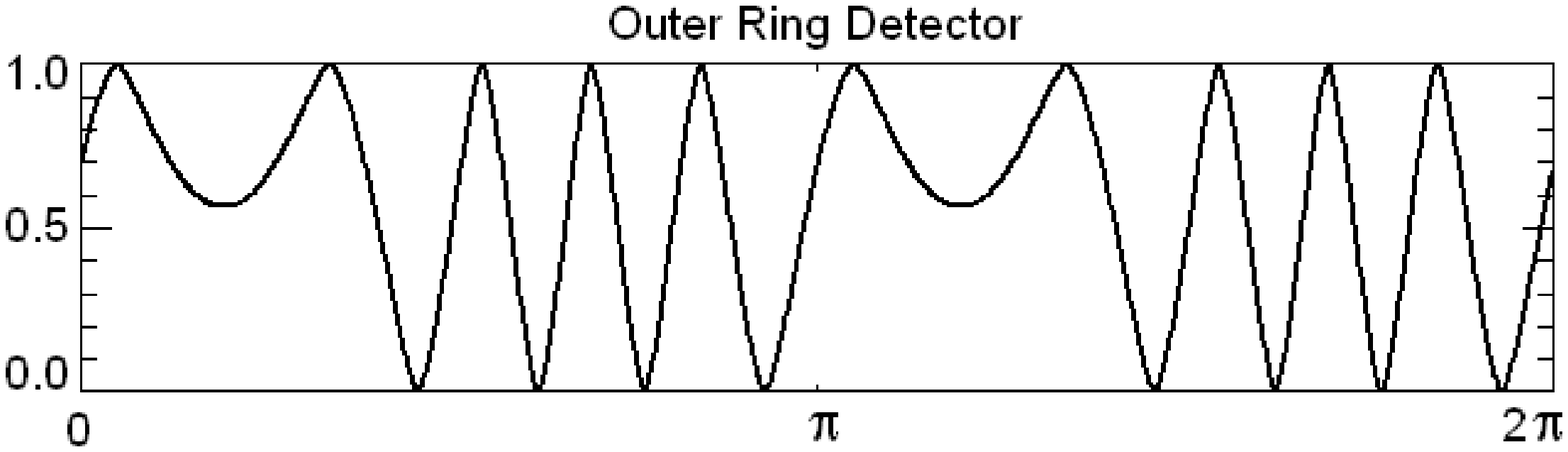} \\
	\includegraphics[width=2.6in]{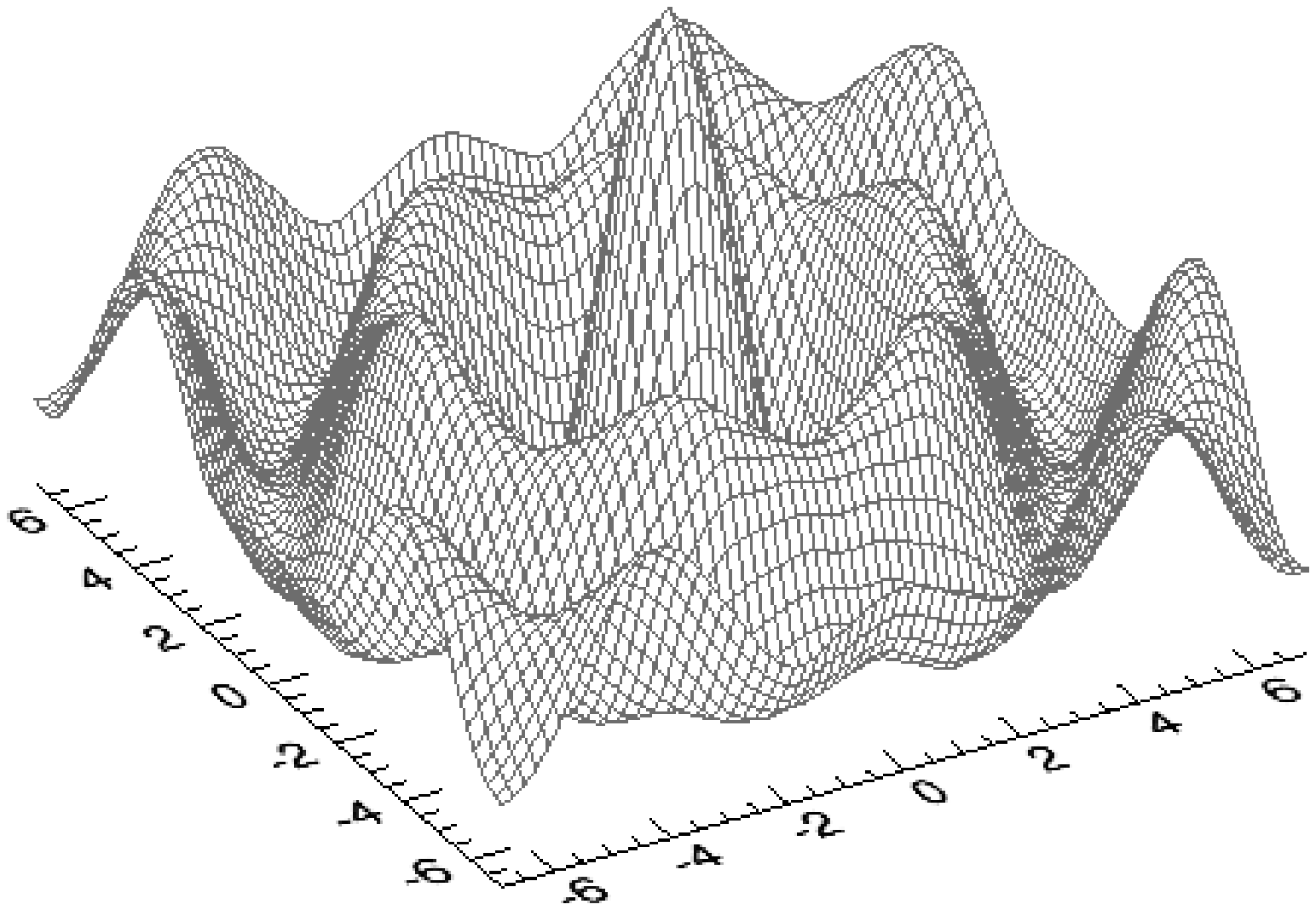}
	\end{center}
	\caption[example]
   { \label{fig:profiles} 
   (Top) Deterministic count profiles for a single point source viewed from three detector positions. The horizontal scale is grid rotation phase, while the vertical scale is unit source intensity. (Bottom) The point-spread function of the prototype LaBRAT RM imager described in the text in sec. \ref{LaBRAT} below, with axes in degrees.}
\end{figure}

In the sections below, we briefly describe the iterative ``Noise Compensating Algebraic Reconstruction'' (NCAR) algorithm that has been developed to deconvolve RM images, the prototype RM imager that has been constructed in our laboratory at Louisiana State University (LSU), and the measurement results that demonstrate the ability of the RM to image sources in the presence of background and to provide super-resolution several times better than the geometric resolution. We conclude by describing the design of an RM telescope suitable for high-sensitivity stand-off gamma-ray measurements or gamma-ray astronomy observations on a high altitude balloon platform.

\section{Image Reconstruction}

The RM grid rotates at a fixed frequency $\omega$. An object scene $S(n)$ produces a time history of counts in each detector $d$,
\begin{equation}
	O_d(t) = \sum_n P_d(n,t)S(n) + N_d(t).
	\label{eq:oRM}
\end{equation}
The instrument response $P_d(n,t)$ is a collection of pre-calculated count profiles for all possible source locations $n$ within the object scene. Fig. \ref{fig:profiles} (top) shows count profile examples of the same source viewed by detectors at three different positions.

To reconstruct the object scene, a cross-correlation image is first generated by the summed convolution of the measured count profiles from each detector with the instrument response:
\begin{equation}
	C(n) = \sum_d \sum_t P_d(n,t)O_d(t) = \sum_{n'} P'(n',n) S(n') + N(n),
	\label{c_full}
\end{equation}
where
\begin{equation}
	P'(n',n) = \sum_d\sum_t P_d(n',t) P_d(n,t).
\end{equation}
The cross-correlation image for a point source reveals the PSF to have a ring pattern centered on the source location (Fig. \ref{fig:profiles}, bottom). Algebraic solutions attempt to solve directly for the object scene by deconvolution of the cross-correlation with the PSF. Direct Demodulation (DDM) \cite{Li1994} is one such technique, which solves the system of equations iteratively while enforcing a positivity condition as a physical constraint on the reconstruction. DDM has been previously shown by its authors to provide super-resolution for RMCs \cite{chen1998}, and we have shown the same ability with RMs in simulation where little or no noise exists \cite{BuddenSPIE2009}. Unlike a statistical technique, however, any algebraic reconstruction method inherently requires that the reconstructed image agrees exactly with the data. Since the data include statistical uncertainty from source and background measurements, this technique may lead to spurious source reconstructions and inaccurate true source locations when the measured signal-to-noise ratio (SNR) is low.

NCAR is an iterative method devised to compensate for the presence of noise in the data \cite{BuddenTNS2010}. The algorithm is based on the DDM technique but includes compensation for noisy data to suppress fluctuations. A modified Gauss-Seidel method is used to solve for the object scene \cite{Barrett1994}. A randomized background layer, derived from Poisson uncertainty in the data, is then added to the cross-correlation image at each iteration. Once the reconstruction converges to the data as well as possible given the noise limitations, subsequent reconstructions are combined in a running average; spurious random peaks are suppressed while true sources remain.
 
\section{Prototype RM Imager}
\label{LaBRAT}

\begin{figure}
	\begin{center}
	\includegraphics[width=3.4in]{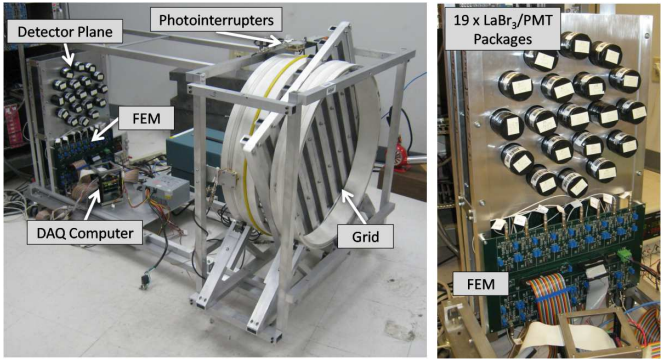}
	\end{center}
	\caption[example]
   { \label{fig:labrat} 
   (Left) Photograph of the LaBRAT instrument. (Right) Closeup of LaBRAT's detection plane.}
\end{figure}

The Lanthanum Bromide-based Rotating Aperture Telescope (LaBRAT, Fig. \ref{fig:labrat}) is a laboratory prototype RM developed at LSU \cite{BuddenIEEE2008}. It was designed to verify the feasibility of the RM technique and the reconstruction algorithm, and as a pathfinder experiment to an eventual high-altitude balloon mission for gamma-ray astronomy in the 30 -- 700 keV energy range. While the RMC has seen limited use in gamma-ray imaging (e.g. \cite{Hurford2002}), the authors are aware of only one other single-grid RM-type instrument; the WATCH experiment \cite{Lund1981, Lund1991}, however, uses a unique and different system of detection than LaBRAT.

The LaBRAT mask is a 2$'$ diameter grid of eight 1.5$''$ $\times$ 24$''$ $\times$ 0.75$''$ thick lead slats, spaced 1.5$''$ apart. The slats are supported by an aluminum structure, and sandwiched between two sections of PVC pipe. The grid sits atop four roller bearings to allow for free rotation. A servo motor is attached by a belt and drives the mask to rotate at approximately 10 rpm. Four photointerrupters are mounted around the grid in order to orient the mask and allow for subsequent folding of the count profiles. 

The detection plane is composed of nineteen Cerium-doped Lanthanum Bromide (Ce:LaBr$_3$) BrilLanCe 380 scintillators manufactured by Saint-Gobain \cite{SaintGobain2010}. LaBr$_3$ provides excellent energy resolution compared to other inorganic scintillators ($< 3$\% FWHM at 662 keV) due primarily to its high light yield. Each detector is hermetically sealed in an aluminum housing with a glass window on one end; this end is optically coupled to a 1.5$''$ Electron Tubes 9102B photomultiplier tube. The detection plane is offset $\sim$1.2 m from the mask, defining a 13.8$^{\circ}$ full-angle field of view and 1.9$^{\circ}$ geometric angular resolution.

The data acquisition system was custom designed and built for LaBRAT. Each PMT anode signal is fed into a separate channel on the data acquisition front end module (FEM), where the signal is split; one of the signals goes to the pre-amplifier circuit and is eventually digitized, while the other is used for triggering. Currently, the system is capable of reading 32 channels, but has been designed with a larger system in mind, suitable for a balloon-flight instrument, and so is easily scaled.

\section{Results}

Monte Carlo simulation results have been reported in \cite{BuddenTNS2010}. Here, two 50 $\mu$Ci $^{137}$Cs sources are used to demonstrate the measured resolving power of LaBRAT. The sources are placed $\sim$10 m from the detection plane. Since the RM was designed primarily as a far-field imager, the reconstruction algorithm assumes photons from a source are mutually parallel when incident on the instrument. Parallax can degrade the image, and so this distance provides a suitable tradeoff between minimal parallax and higher measured source rate. With a 5 Hz per source count rate and 30 Hz background in the selected energy range, exposure times of 5 -- 10 hours are used for these measurements.

\begin{figure}
	\begin{center}
	\includegraphics[width=3.0in]{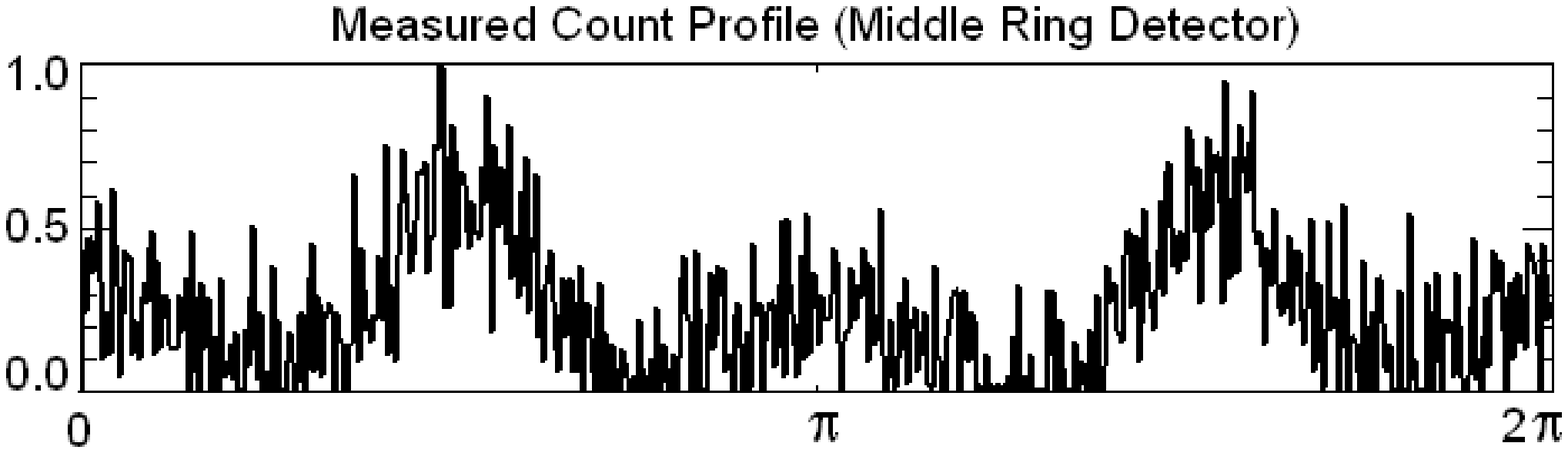} \\
	\includegraphics[width=3.0in]{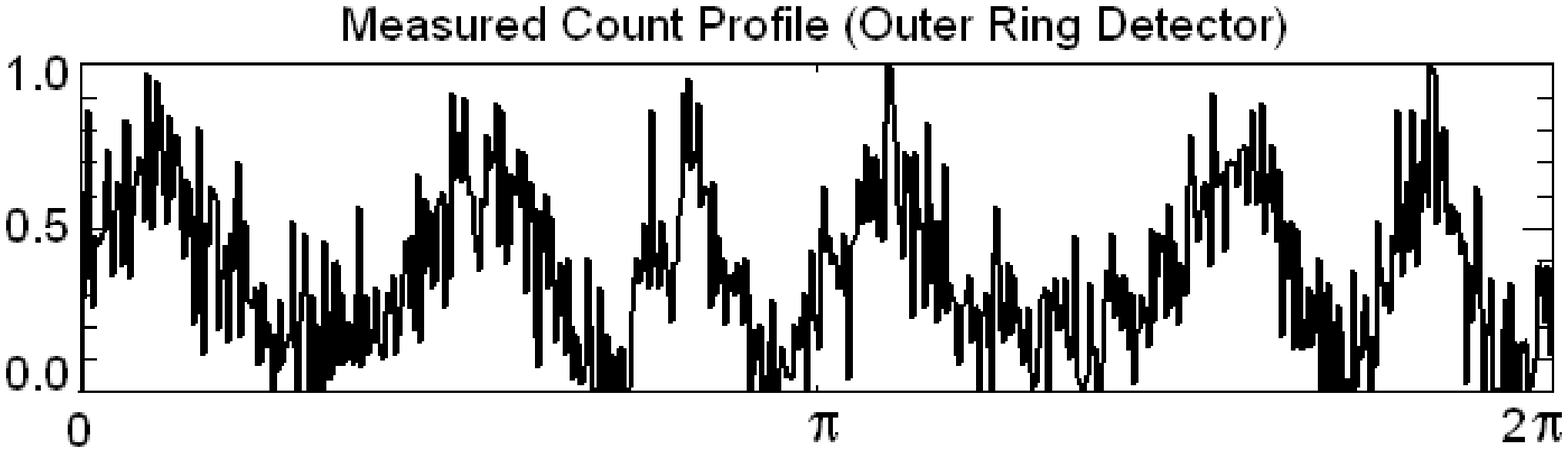} \\
	\end{center}
	\caption[example]
   { \label{fig:cnt_profile} 
   Normalized measured count profiles for a middle and outer-ring detector for one of the observations shown in Figure \ref{fig:results}.}
\end{figure}

\begin{figure}
	\begin{center}
	\begin{tabular}{c c}
		2$^{\circ}$22$'$ & 1$^{\circ}$46$'$ \\
		\includegraphics[width=1.5in]{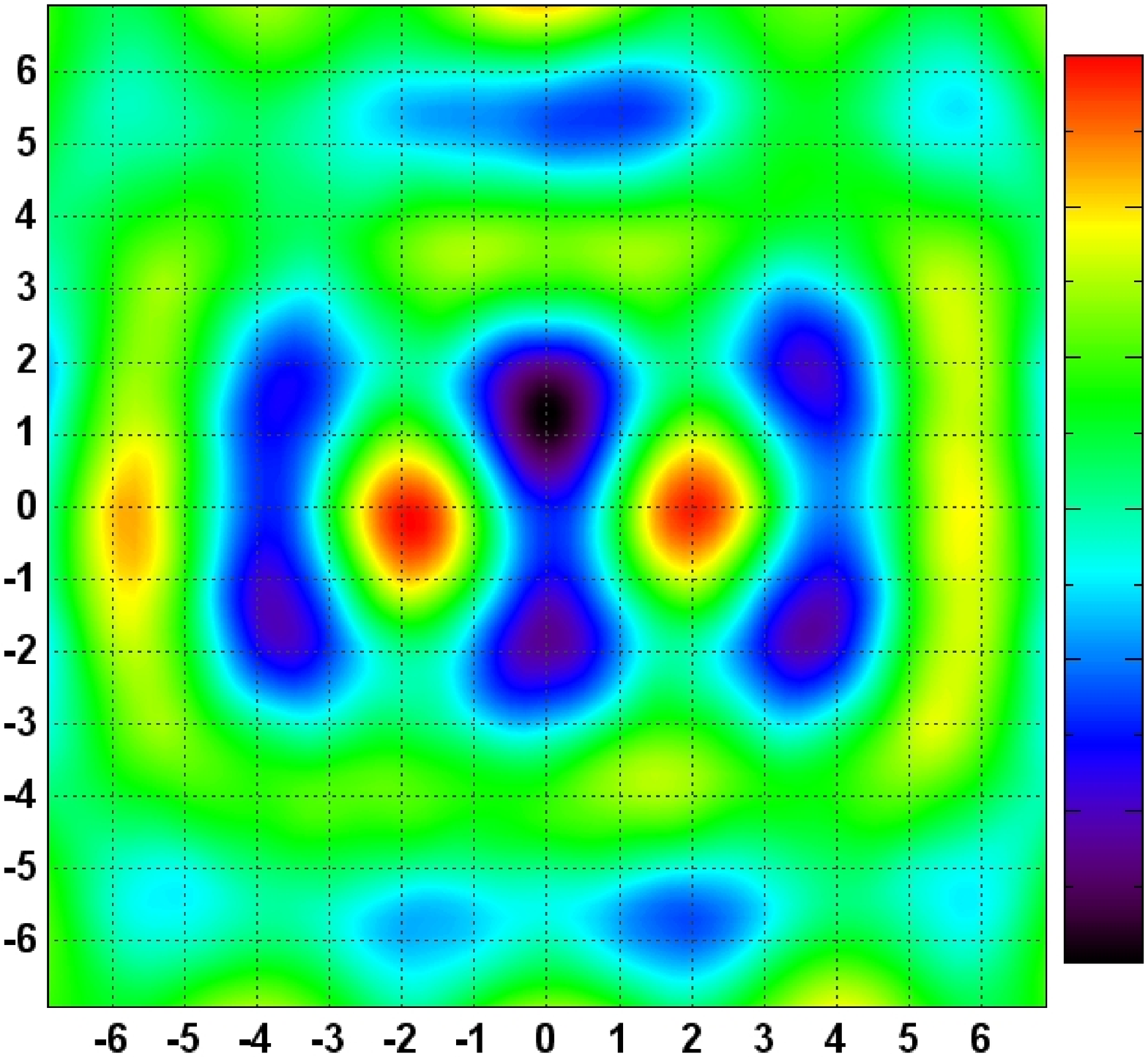} & \includegraphics[width=1.5in]{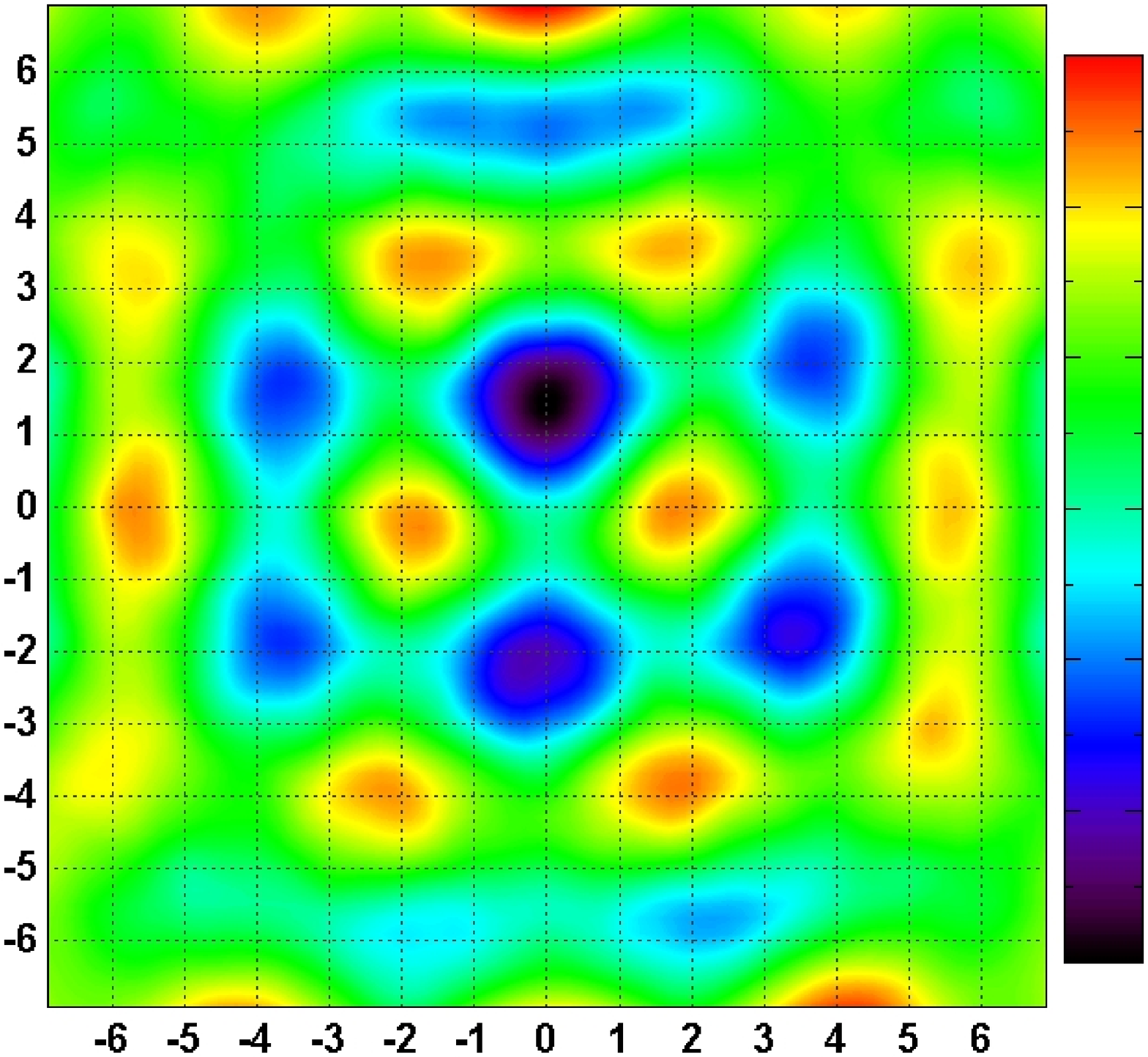} \\
		\includegraphics[width=1.5in]{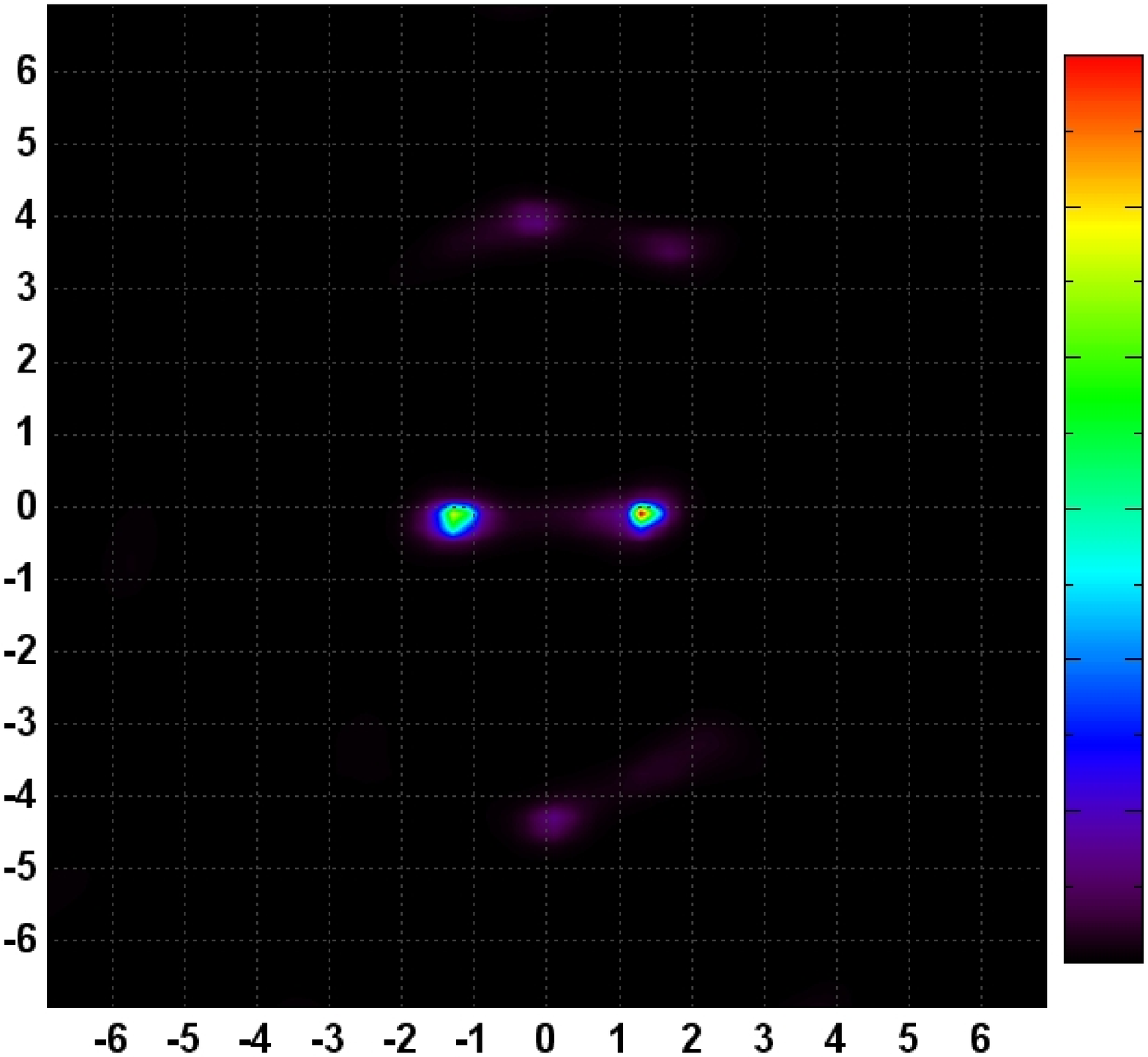} & \includegraphics[width=1.5in]{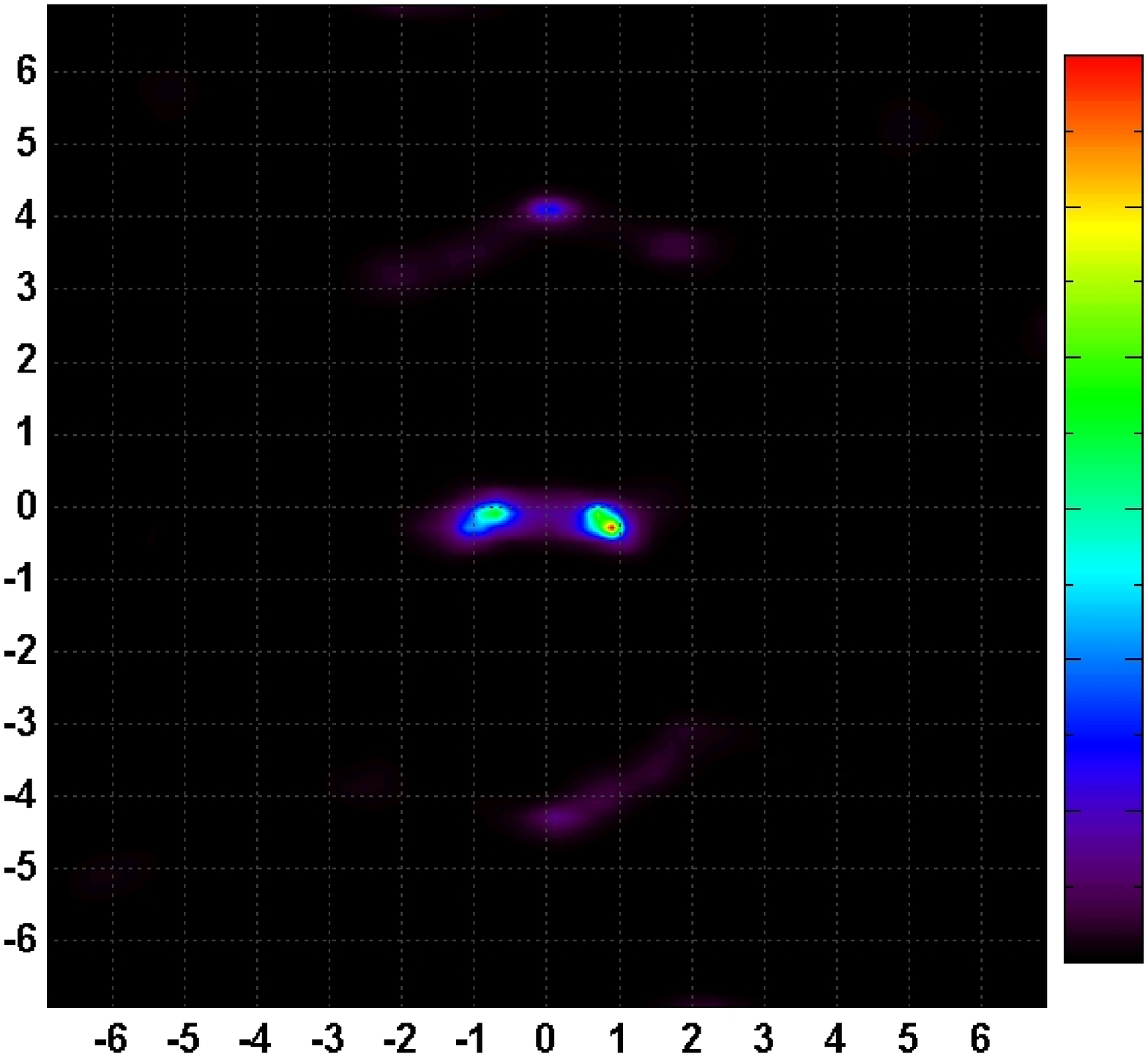} \\
		\includegraphics[width=1.5in]{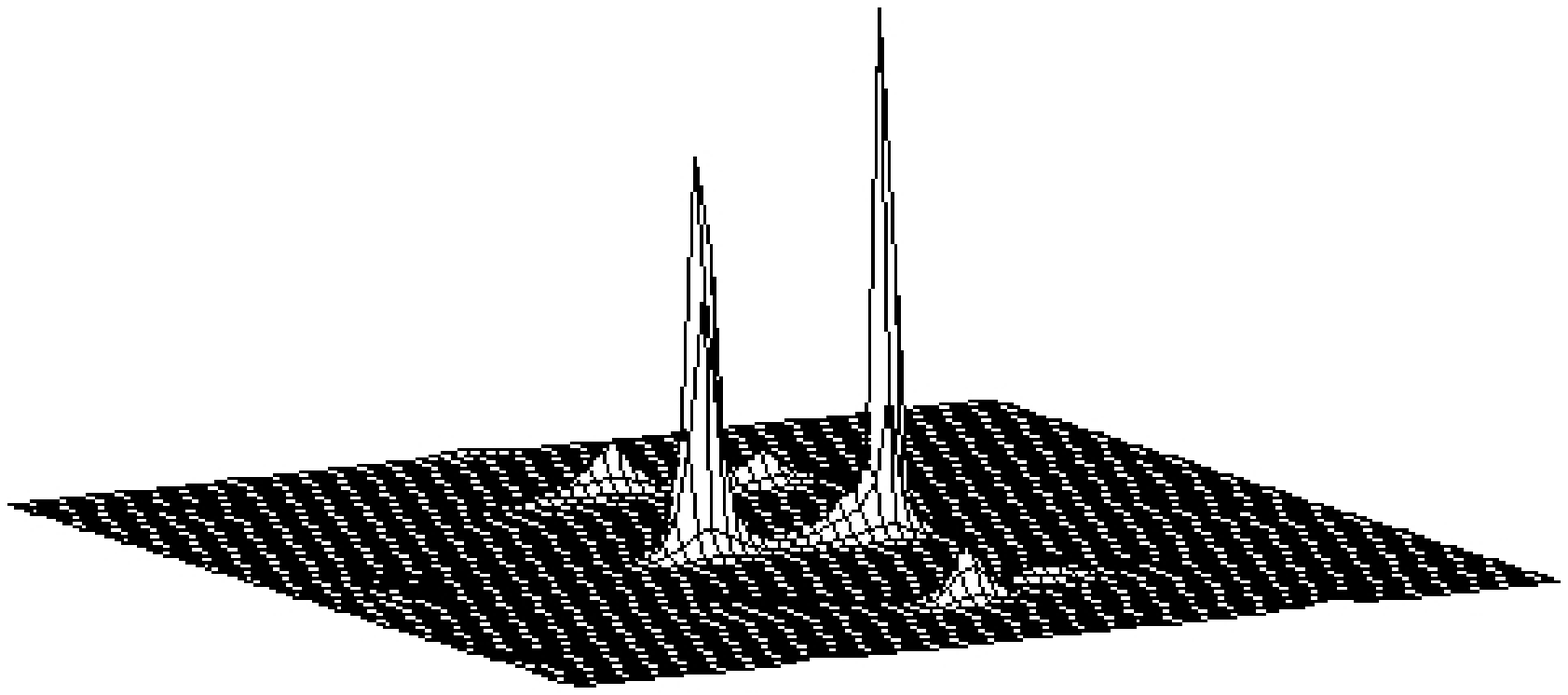} & \includegraphics[width=1.5in]{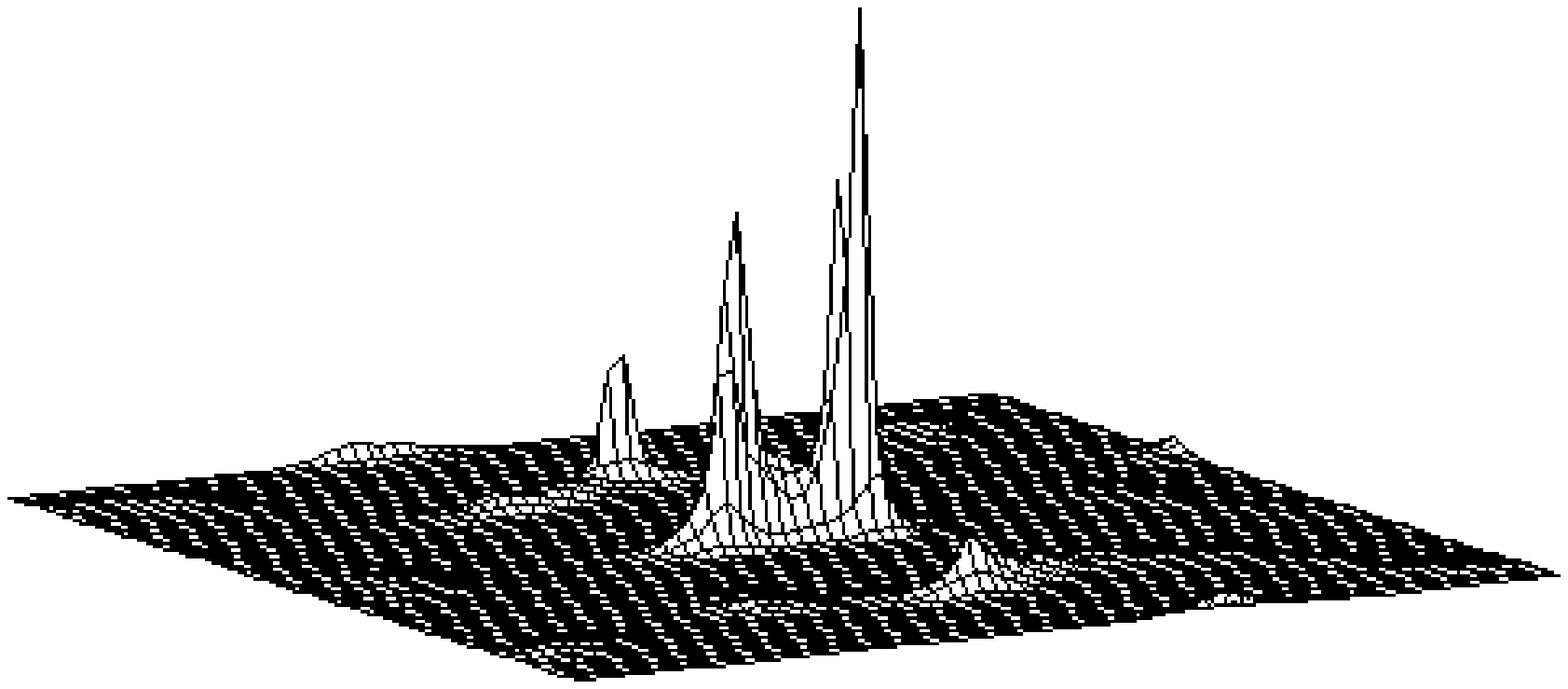} \\
		& \\
		1$^{\circ}$10$'$ & 35$'$ \\
		\includegraphics[width=1.5in]{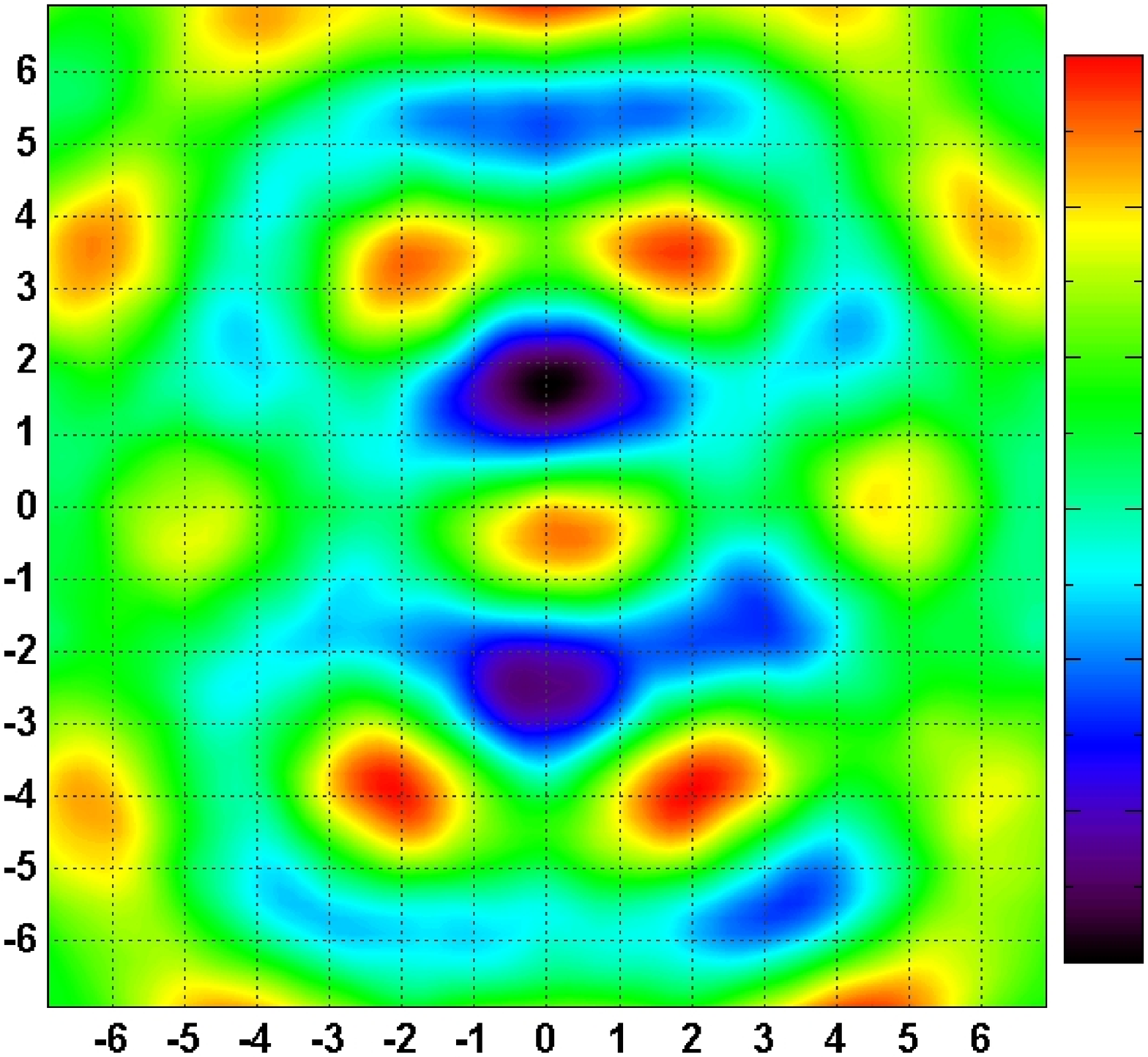} & \includegraphics[width=1.5in]{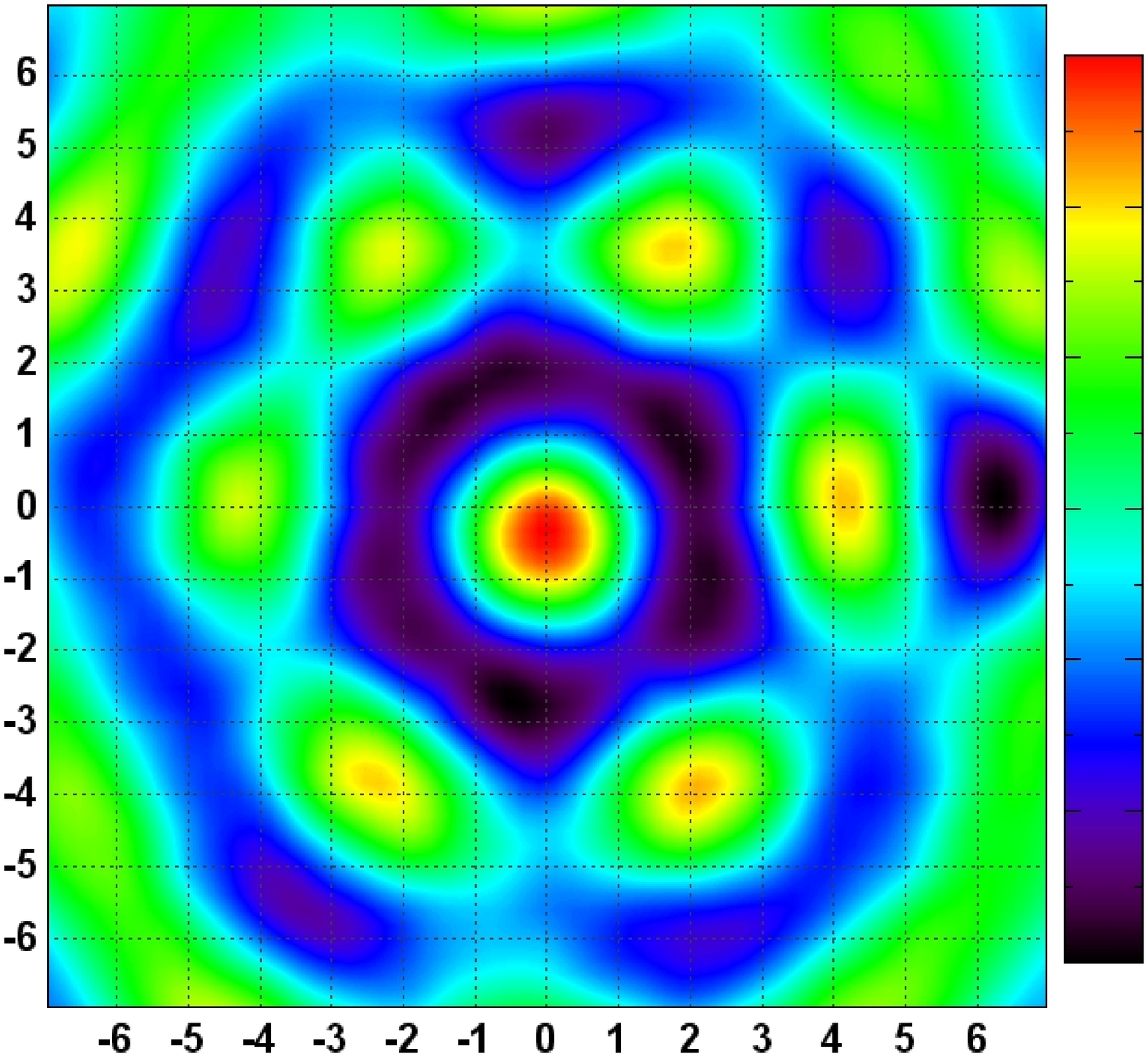} \\
		\includegraphics[width=1.5in]{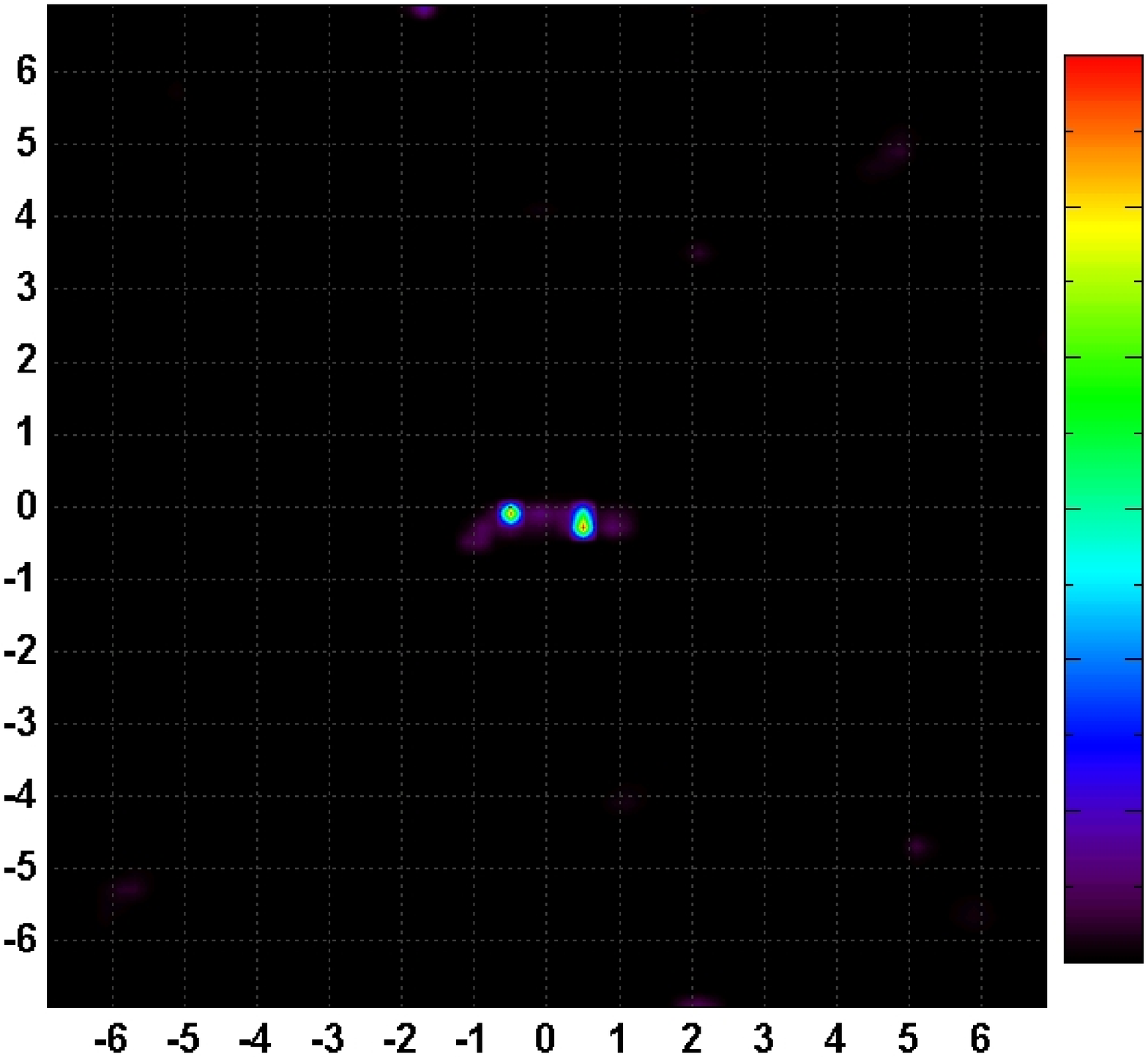} & \includegraphics[width=1.5in]{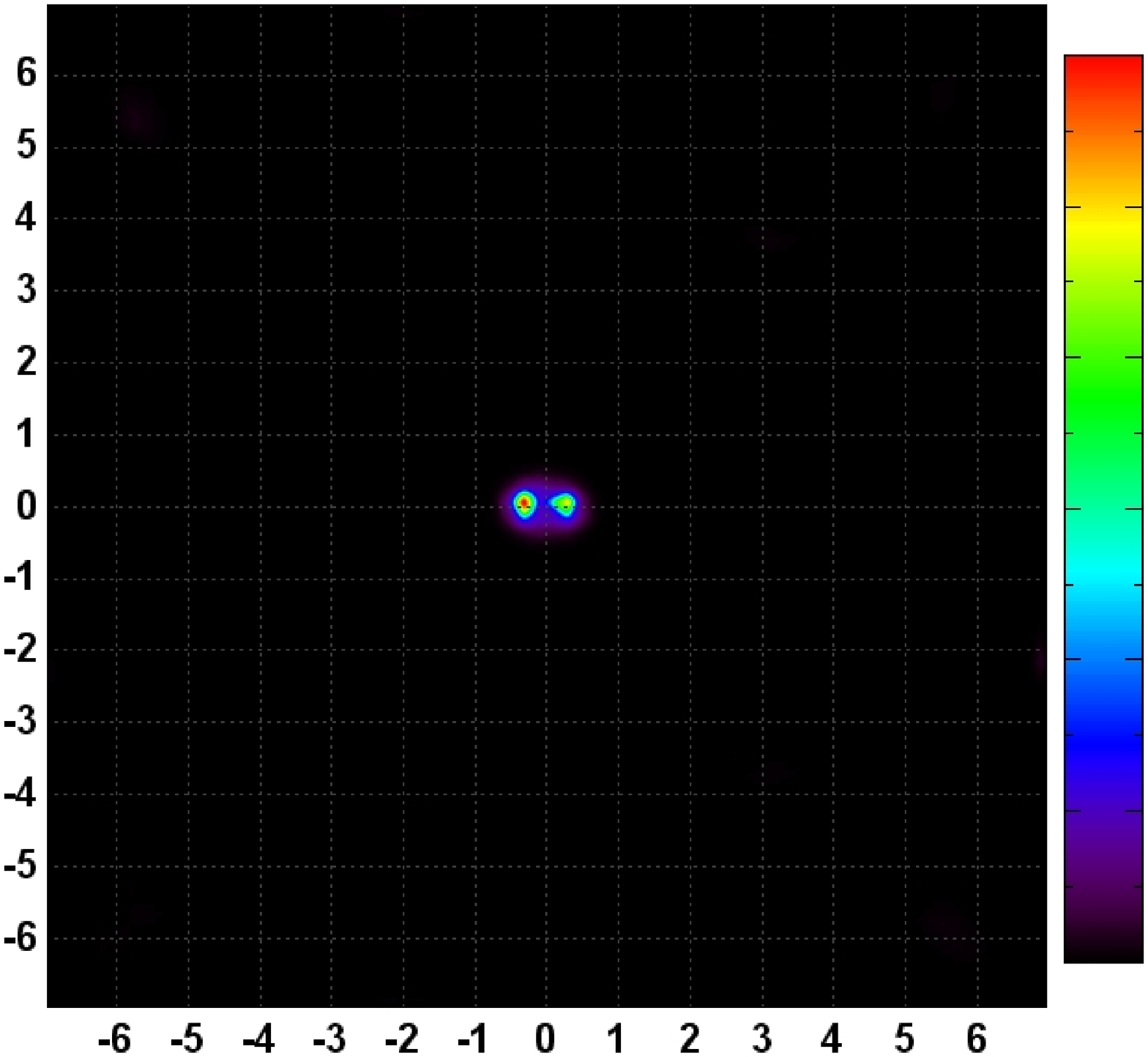} \\
		\includegraphics[width=1.5in]{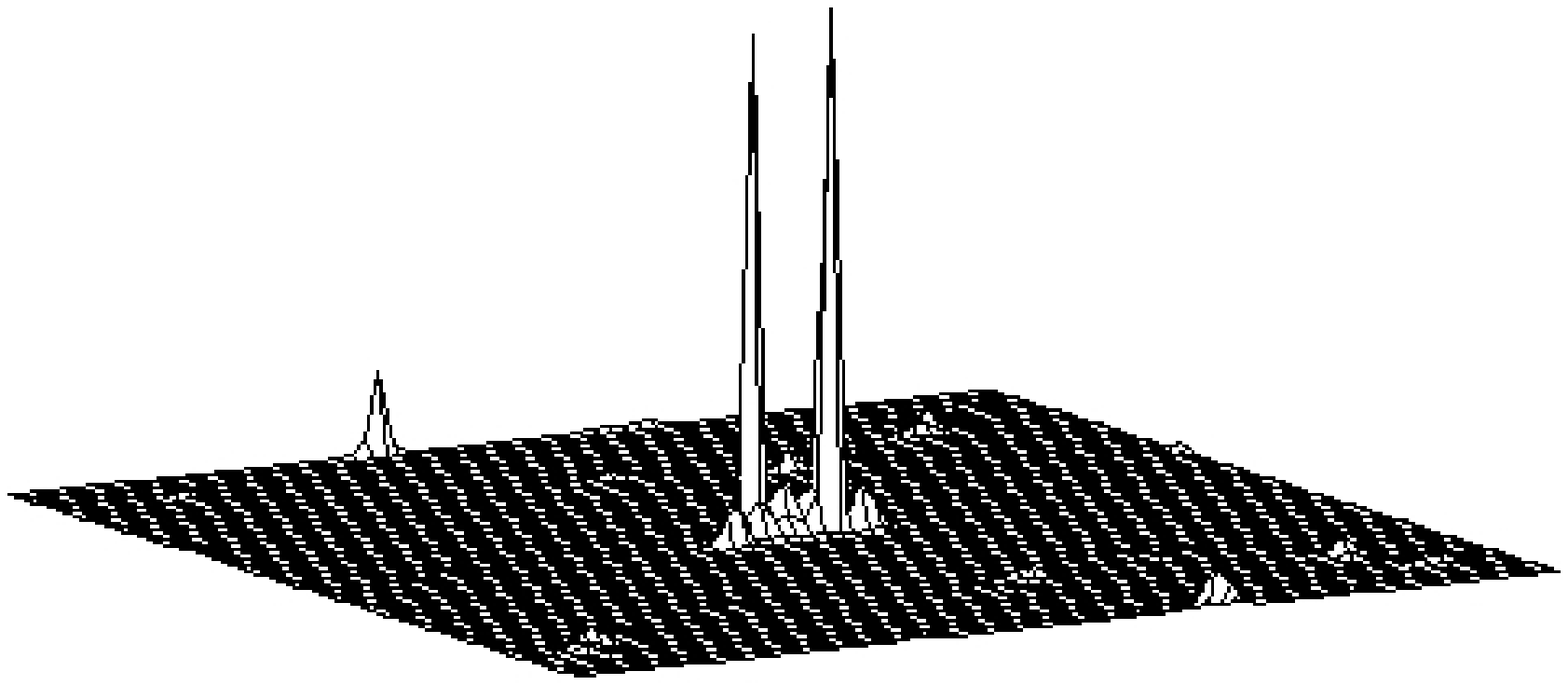} & \includegraphics[width=1.5in]{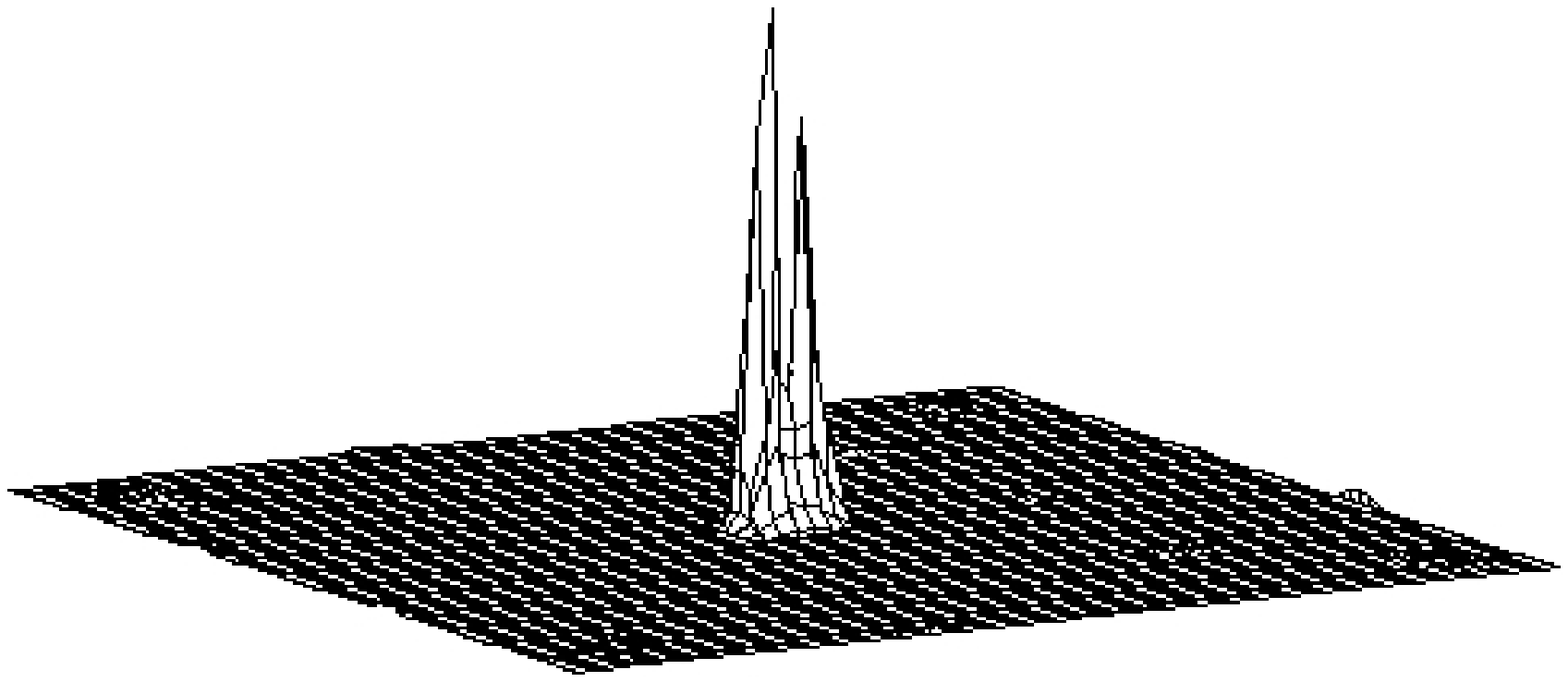}
	\end{tabular}
	\end{center}
	\caption[example]
   { \label{fig:results} 
   Contour plot of cross-correlation image (upper row), contour plot of final NCAR reconstruction (middle row), and surface mesh of final reconstruction (bottom row) for two sources at four different angular separations, as imaged  by LaBRAT. Note, in particular, the cross-correlation images for 1$^{\circ}$10$'$ and 35$'$ separation, which are not capable of resolving the two sources independently; the final reconstructions using the NCAR algorithm (middle and lower images), however, are capable of achieving this resolving power. Axes in degrees.}
\end{figure}

The sources are placed at varying angular separations and imaged (Fig. \ref{fig:results}). The 0.75$''$ thick mask only attenuates $\sim 90\%$ of the 662 keV photons observed, and so an absorption term is included in the calculation of the count rate profile $P_d(t)$. At 2$^{\circ}$22$'$, the cross-correlation image shows the two sources, and the reconstruction successfully removes the sidelobes. At 1$^{\circ}$46$'$, almost equal to the geometric resolution of the instrument, the cross-correlation image begins to deteriorate. At even smaller angular separations (1$^{\circ}$10$'$ and 35$'$), only a single central peak is visible in the cross-correlations. In all four cases, however, the final reconstructions derived from the NCAR algorithm are able to deconvolve the data and fully resolve the two sources down to a separation of at least 35$'$, successfully achieving super-resolution of a factor greater than 3. As additional improvements are made to the prototype, resolving power of 15 -- 20$'$ is expected, as has been shown in simulation \cite{BuddenSPIE2009}.

\section{Discussion and Conclusions}

Coupled with an appropriate reconstruction technique, it appears that a rotational modulator is a suitable alternative to a coded aperture. It maintains the same throughput, and is capable of achieving comparable angular resolution with a far simpler and cheaper detection plane. We have built and tested a laboratory prototype, LaBRAT, to demonstrate this concept, and have shown the ability to achieve angular resolution $\sim 3\times$ better than the intrinsic geometric angular resolution defined by the instrument. 

Calibrations of the newly-built data acquisition system are still ongoing. Based on earlier simulations and results with a previous version of the data acquisition system \cite{BuddenIEEE2009}, we expect to be able to separate sources at 15 -- 20$'$. Further improvement is expected by dithering the telescope around the source position. (For a balloon-borne version of the instrument used for gamma-ray astronomy, as described below, the motion of the balloon platform will provide a natural dithering motion.) A detailed discussion of the effects of dithering, the applications to moving sources and moving detectors, and angular resolution results as a function of signal-to-background rates will be published in a separate report.

The High Altitude Rotational Modulator for Energetic radiation Imaging (HARMEnI) gamma-ray astronomy telescope design is based on the LaBRAT prototype results and the NCAR imaging algorithm. HARMEnI will serve as a pathfinder mission for a satellite-borne instrument to survey and monitor black-hole candidates, observing at energies between 30 and 700 keV. Relative to LaBRAT, the initial HARMEnI telescope will have increased sensitivity and field of view. Ten lead bars comprise the mask, and the detection plane is a tightly-packed arrangement of 37 LaBr$_3$ crystals -- the same as those used with LaBRAT. HARMEnI provides a field of view of 20$^{\circ}$ and geometric angular resolution of 1.9$^{\circ}$. Since HARMEnI's geometric resolution is the same as that of the LaBRAT prototype, an ultimate resolving power of 15 -- 20$'$ is anticipated.

Background rates at balloon float altitude have been measured with a similar Ce:LaBr$_3$ crystal flown on a long duration (17 day) Antarctic balloon flight \cite{LaxatevIEEE2008}. The main contributors to the background are the atmospheric and cosmic diffuse flux and the internal background due to $^{138}$La decays. The expected background rates for HARMEnI are estimated using the measured data from the Antarctic flight and assuming a  5 mm thick passive graded lead-tin-copper shield extends from the detection plane to the mask and covers the back of the detectors. Based on the expected rates, the observation time necessary for the Crab Nebula will be significantly less than one hour.






\section*{Acknowledgement}

This work has been supported in part by US DOE NNSA Cooperative Agreement DE-FC52-04-NA25683.  B. Budden thanks the Louisiana Board of Regents, under agreement NASA/LEQSF(2005-2010)-LaSPACE and NASA/LaSPACE under grant NNG05GH22H, and the Coates Foundation for support during this project.







\end{document}